\documentclass[onecolumn,showpacs,amsmath,amssymb,floatfix]{revtex4}

\usepackage{longtable,array}
\usepackage{graphicx}

\usepackage{bm}

\begin{document}

\newcommand{\half}{\mbox{$\textstyle \frac{1}{2}$}}
\newcommand{\ket}[1]{\left | \, #1 \right \rangle}
\newcommand{\bra}[1]{\left \langle #1 \, \right |}
\newcommand{\beq}{\begin{equation}}
\newcommand{\eeq}{\end{equation}}
\newcommand{\bea}{\begin{eqnarray}}
\newcommand{\eea}{\end{eqnarray}}
\newcommand{\req}[1]{Eq.\ (\ref{#1})}
\newcommand{\gcc}{{\rm~g\,cm}^{-3}}
\newcommand{\Compton}{\lambda\hspace{-.44em}\raisebox{.6ex}{\mbox{-$\!$-}}%
\raisebox{-.3ex}{}_{\hspace{-1pt}{\mbox{$_\mathrm{C}$}}}}
\newcommand{\kB}{k_\mathrm{B}}
\newcommand{\omc}{\omega_\mathrm{c}}
\newcommand{\omg}{\omega_\mathrm{g}}
\newcommand{\mel}{m_e}
\newcommand{\xr}{x_{\rm r}}
\newcommand{\EF}{\epsilon_{\rm F}}
\newcommand{\Ne}{{\cal N}_B(\epsilon)}
\newcommand{\Necl}{{\cal N}_0(\epsilon)}
\newcommand{\dfde}{{\partial f^{(0)} \over\partial\epsilon}}
\newcommand{\am}{a_\mathrm{m}}
\newcommand{\dd}{{\rm\,d}}
\newcommand{\vB}{\bm{B}}
\newcommand{\dotZ}{\mbox{$\dot{\mbox{Z}}$}}
\newcommand{\msun}{\mbox{$M_\odot$}}

\def\la{\;\raise0.3ex\hbox{$<$\kern-0.75em\raise-1.1ex\hbox{$\sim$}}\;}
\def\ga{\;\raise0.3ex\hbox{$>$\kern-0.75em\raise-1.1ex\hbox{$\sim$}}\;}
\def\lesssim{\;\raise0.3ex\hbox{$<$\kern-0.75em\raise-1.1ex\hbox{$\sim$}}\;}
\def\gtrsim{\;\raise0.3ex\hbox{$>$\kern-0.75em\raise-1.1ex\hbox{$\sim$}}\;}

\title{Astrophysical S-factors for fusion reactions involving
C, O, Ne and Mg isotopes}

\author{M. Beard$^1$, A. V. Afanasjev$^2$, L. C. Chamon$^3$,\\
 L. R. Gasques$^4$, M. Wiescher$^1$, and D. G.
Yakovlev$^5$}
\affiliation{$^1$ Department of Physics $\&$ The Joint Institute
for Nuclear Astrophysics, University of Notre Dame,  Notre Dame,
IN 46556 USA. \\
$^2$ Department of Physics and Astronomy, Mississippi State University,
P.O.~Drawer 5167, Mississippi 39762-5167, USA. \\
$^3$ Departamento de F\'{\i}sica Nuclear, Instituto de F\'{\i}sica da
Universidade de S\~ao Paulo, Caixa Postal 66318, 05315-970, S\~ao Paulo, SP,
Brazil. \\
$^4$ Centro de F\'{i}sica Nuclear da Universidade de Lisboa,
Av.\ Prof.\ Gama Pinto 2, 1649-003 Lisboa, Portugal. \\
$^5$ Ioffe Physical Technical Institute, Poliekhnicheskaya 26, 194021
St.-Petersburg, Russia.}

\date{\today}
\begin{abstract}
Using the S\~ao Paulo potential and the barrier penetration
formalism we have calculated the astrophysical factor $S(E)$ for 946
fusion reactions involving stable and neutron-rich isotopes of C, O,
Ne, and Mg for center-of-mass energies $E$ varying from 2 MeV to
$\approx$18--30 MeV (covering the range below and above the Coulomb
barrier). We have parameterized the energy dependence $S(E)$ by an
accurate universal 9-parameter analytic expression and present
tables of fit parameters for all the reactions. We also discuss the
reduced 3-parameter version of our fit which is highly accurate at
energies below the Coulomb barrier, and outline the procedure for
calculating the reaction rates. The results can be easily converted
to thermonuclear or pycnonuclear reaction rates to simulate various
nuclear burning phenomena, in particular, stellar burning at high
temperatures and nucleosynthesis in high density environments.
\end{abstract}


\maketitle

\section{Introduction}
\label{introduction}

\indent Nuclear reactions are extensively studied theoretically
and in laboratory experiments. They are important for modelling
many physical phenomena occurring in astrophysical environments,
such as energy production and nucleosynthesis in stars at all
stages of their evolution (e.g., Refs.\
\cite{bbfh57,fh64,clayton83}). Nuclear reactions are important for
the structure and evolution of main sequence and red giant stars,
during the late pre-supernova phase of massive stars and during
core-collapse supernova explosions.

Moreover, nuclear burning is also important in high-density stellar
environments of white dwarfs and neutron stars which are compact
stars at the final stage of their evolutionary development
\cite{st83}. The burning drives nuclear explosions in surface layers
of accreting white dwarfs (nova events), in cores of massive
accreting white dwarfs (type Ia supernovae)
\cite{NiWo97,hoeflich06}, and in surface layers of accreting neutron
stars (type I X-ray bursts and superbursts; e.g., Refs.\
\cite{sb06,schatz03,cummingetal05,Brown06}). The nova events and
type I X-ray bursts are mostly produced by the burning of hydrogen
in the thermonuclear regime, without any strong effect of plasma
screening on Coulomb tunneling of the reacting nuclei. Type Ia
supernovae and superbursts are driven by the burning of carbon,
oxygen, and heavier elements
(e.g., Refs.\ \cite{sb06,schatz03,cummingetal05,Brown06}) at high densities,
where the plasma screening effect can be substantial. It is likely
that pycnonuclear burning of neutron-rich nuclei (e.g.,
$^{34}$Ne+$^{34}$Ne) in the inner crust of accreting neutron stars
in X-ray transients \cite{hz90,hz03,Brown06} (in binaries with
low-mass companions) provides an internal heat source for these
stars. If so, it powers \cite{Brown98} thermal surface X-ray
emission of neutron stars observed in quiescent states of X-ray
transients (see, e.g., Refs.\ \cite{Brown06,pgw06,lh07}).

Nuclear fusion occurs in a wide range of temperatures and densities
of stellar matter (at densities $\rho \lesssim 10^{10}$ g~cm$^{-3}$
in white dwarfs and at $\rho \lesssim 10^{13}$ g~cm$^{-3}$ in
neutron stars). It can proceed in five regimes as described by
Salpeter and Van Horn \cite{svh69} (also see
\cite{gas2005,yak2006,cdi07,cd09} and references therein). These are
two thermonuclear regimes (with weak and strong plasma screening),
two pycnonuclear regimes (due to zero-point vibrations of atomic
nuclei in crystalline lattice, with and without thermal effects in
nucleus motion), and the intermediate thermo-pycnonuclear regime.
The burning may involve many different, stable as well as very
neutron-rich and unstable nuclei. Neutron-rich nuclei are especially
important in deep neutron star crust, where they can be stabilized
against beta-decay by the presence of the Fermi sea of highly
energetic degenerate electrons \cite{st83,hpy07}.

The nuclear fusion rates depend directly on the reaction cross
section which can be expressed in terms of the astrophysical factor
$S(E)$, with $E$ as the center-of-mass energy of the interacting
nuclei. Stellar burning often proceeds at low temperatures with
sub-picobarn reaction cross sections that are usually not accessible
by direct measurements in laboratory experiments. Therefore, it is
necessary to calculate $S(E)$ for the reactions of interest by a
reliable theoretical method in the important energy range and
approximate the data by analytical expressions for rapid conversion
into thermonuclear or pycnonuclear reaction rates as outlined in our
previous work \cite{yak2006}. In this paper, we calculate (Sec.\
\ref{theory}) the astrophysical factors $S(E)$ for 946 fusion
reactions involving different isotopes of C, N, O, Ne, and Mg
(between the valley of stability and the neutron drip line) using
the barrier penetration model and the S\~ao Paulo
potential~\cite{saoPauloTool}. We approximate our results (Sec.\
\ref{fits}) by analytic expressions convenient for real time
applications. A numerical example is discussed in Sec.\
\ref{example}; calculation of the reaction rates is outlined in
Sec.\ \ref{s:rates}, and we conclude in Sec.\ \ref{s:concl}.

\section{Calculation of astrophysical S-factors}
\label{theory}

\indent The cross section $\sigma(E)$ for a fusion reaction of two
nuclei,
\begin{equation}
         (A_1,Z_1) + (A_2,Z_2),
\label{react}
\end{equation}
at the center-of-mass energy $E$ can be expressed in terms of the
astrophysical factor $S(E)$ by
\begin{equation}
   \sigma(E) = {1\over E}\, \exp(-2 \pi \eta)\,S(E),
\label{sigma}
\end{equation}
where $\eta= Z_1Z_2e^2/(\hbar v)$ is the Sommerfeld parameter,
$v=\sqrt{2E/ \mu}$ the relative velocity of reacting nuclei at large
separations, and $\mu$ the reduced mass. The factor $\exp(-2 \pi
\eta)$ comes from the probability of penetration through the Coulomb
barrier $V_C(r)=Z_1Z_2e^2/r$ with zero orbital angular momentum,
$\ell=0$, assuming that the barrier extends to $r \to 0$ (for
point-like nuclei); $1/ E$ factorizes out the well-known
pre-exponential low-energy dependence of $\sigma(E)$. The advantage
of this approach is that $S(E)$ is a much more slowly varying
function of $E$ than $\exp(-2 \pi \eta)$ and $\sigma(E)$. It is
easier to extrapolate $S(E)$ to low energies $E$ of astrophysical
interest, than $\sigma(E)$.

We calculate $S(E)$ for a number of fusion reactions using the S\~ao
Paulo potential in the context of the barrier penetration model as
outlined in previous studies \cite{gas2005,yak2006,saoPauloTool}.
The S\~ao Paulo potential is a parameter-free model for the real
part of the nuclear interaction. Underpinning this potential are the
contributions from nonlocality due to quantum effects arising from
the Pauli exclusion principle
\cite{chamon2002,candido1997,chamon1997,chamon1998}.  The
interaction takes into account the exchange of nucleons between
reacting nuclei. The potential has been extensively tested in the
description of different nuclear interaction processes (such as
quasi-elastic scattering and fusion reactions), over a broad
spectrum of energies
\cite{gas2004,chamon2004,chamon2002,alvarez2003,alvarez1999,sil2001,
gas2002,rossi2002,gas2003,gascham2003,saoPauloTool}. In the context
of the S\~ao Paulo potential, the  real part of the nuclear
interaction is associated with the folding potential through the
relation
\begin{equation}
   V_\mathit{SP}(r,E)=V_F(r)\, \exp( - 4 \mathrm{v}^2/c^2),
\label{sp}
\end{equation}
where $r$ is a distance between the centers of the reactants, and
$c$ is the speed of light. The exponential term houses the effects
of Pauli nonlocality. The relative velocity $\mathrm{v}(r,E)$ of the
nuclei in the given model at a separation $r$ is defined as
\begin{equation}
     \mathrm{v}^2(r,E) = (2/\mu) \,
    \left[ E-V_C(r)-V_\mathit{SP}(r,E) \right] \;,
\label{speed}
\end{equation}
where $V_C(r)$ is the Coulomb potential.

The folding potential used in Eq.\ (\ref{sp}) has a field strength
of $V_0 = -456$ Mev fm$^3$; it is given by
\begin{equation}
   V_F(r) = \int \rho_1(\bm{r}_1) \; \rho_2(\bm{r}_2) \;
   V_{0} \; \delta(\bm{r}-\bm{r}_{1}+\bm{r}_{2}) \; d\bm{r}_1 \; d\bm{r}_2 .
\label{fold}
\end{equation}
It is convoluted over the 
nuclear 
matter densities $\rho_1(\bm{r}_1)$ and $\rho_2(\bm{r}_2)$
of the nuclei involved in the reaction. This technique is called the
zero-range approach for the folding potential. It is equivalent
\cite{chamon2002} to adopting the M3Y effective nucleon-nucleon
interaction with nucleon densities of nuclei.

The barrier penetration model calculates the reaction cross section
$\sigma(E)$ using the standard partial wave ($\ell=0,1,\ldots$)
decomposition and the effective potential $V_\mathrm{eff}(r,E)$. The
latter is constructed from the contributions of the Coulomb, nuclear
and centrifugal potentials,
\begin{equation}
     V_\mathrm{eff}(r,E)= V_C(r)+V_\mathit{SP}(r,E)+
     \frac{ \hbar^2 \ell (\ell+1)}{2 \mu r^2}.
\label{veff}
\end{equation}
At low energies of our interest, the main contribution to
$\sigma(E)$ comes from the $\ell=0$ (s-wave) channel. Once
$\sigma(E)$ is calculated, we determine $S(E)$ from Eq.\
(\ref{sigma}).

The systematics for the 
matter densities $\rho_1(\bm{r}_1)$ and
$\rho_2(\bm{r}_2)$ obtained with a two-parameter Fermi (2pF)
\cite{chamon2002} shape might not be appropriate to provide the
nuclear potential for reactions involving neutron-rich nuclei. The
relativistic Hatree-Bogoliubov approach \cite{rhb1,rhb3} has been
shown to perform well in describing nuclear properties across the
nuclear chart, including the region of neutron-rich nuclei
\cite{rhb1,rhb4,rhb5}. In the present manuscript, the density
distributions of all nuclei were obtained within this approach,
employing the NL3 \cite{rhb4} parametrization of the relativistic
mean field Lagrangian. For that reason, the values of $S(E)$
computed for some reactions ($^{12}$C+$^{12}$C, $^{12}$C+$^{16}$O,
$^{16}$O+$^{16}$O), which we considered previously
\cite{gas2005,yak2006} using the 2pF model, are now somewhat
different.

We have already used \cite{gas2004,saoPauloTool} the S\~ao Paulo
potential with the NL3 nucleon density distribution in the context
of the barrier penetration model and calculated $S(E)$ for a
number of reactions involving stable and neutron-rich nuclei. The
results were compared with experimental data and with theoretical
results calculated by other models such as coupled-channels and
fermionic molecular dynamics ones. As detailed in
\cite{saoPauloTool}, the S\~ao Paulo potential gives reasonably
accurate $S(E)$ for non-resonant nuclear reactions; the method is
parameter-free and relatively simple for generating a set of data
for many reactions involving different isotopes.

Here we calculate astrophysical $S$-factors for 946 fusion reactions
involving combinations of carbon, oxygen, neon, and magnesium
isotopes as summarized in Table \ref{tab:reactions}. We consider 10
reaction types, such as C+C and O+Ne, with the range of mass numbers
for both species given in Table \ref{tab:reactions}. For each
reaction, we compute $S(E)$ in the energy range from 2~MeV to a
maximum value $E_\mathrm{max}$ (also given in Table
\ref{tab:reactions}) in energy steps of 0.1~MeV. The value of
$E_\mathrm{max}$ was chosen in such a way that wide energies ranges
below and above the Coulomb barrier are covered. The fifth column in
Table \ref{tab:reactions} presents the number of considered
reactions of a given type and the last column refers to a table
which lists fit parameters of analytic approximation to $S(E)$ for
such reactions (Sec.\ \ref{fits}).

\begin{table}
\caption[]{Fusion reactions $(A_1,Z_1)+(A_2,Z_2)$ under consideration}
\label{tab:reactions}
\begin{center}
\begin{tabular}{c c c c c c c}
\hline \hline
Reaction & $\quad A_1 \quad $ & $\quad A_2 \quad $ &
~$E_\mathrm{max}$~ & Nr.\ of & ~Table~ \\
type   & even & even &  MeV & cases & of fits \\
\hline
C + C & 10--24 & 10--24 & 17.9 & 36 & \ref{tab:cc} \\
C + O & 10--24 & 12--28 & 17.9 & 72 & \ref{tab:co} \\
C+Ne  & 10--24 & 18--40 & 19.9 & 96 & \ref{tab:cne} \\
C+Mg  & 10--24 & 20--46 & 19.9 & 112 & \ref{tab:cmg} \\
O + O & 12--28 & 12--28 & 19.9 & 45 & \ref{tab:oo} \\
O+Ne  & 12--28 & 18--40 & 21.9 & 108 & \ref{tab:one} \\
O+Mg  & 12--28 & 18--46 & 21.9 & 126 & \ref{tab:omg} \\
Ne+Ne & 18--40 & 18--40 & 21.9 & 78 & \ref{tab:nene} \\
Ne+Mg & 18--40 & 20--46 & 24.9 & 168 & \ref{tab:nemg} \\
Mg+Mg & 20--46 & 20--46 & 29.9 & 105 & \ref{tab:mgmg} \\
\hline \hline
\end{tabular}
\end{center}
\end{table}

In this study we focus on systems of even-even nuclei. Due to
pairing, these nuclei are expected to be more stable in the
astrophysical burning environments in question, for instance, in an
accreted crust of a neutron star \cite{H&Z1989}.

The quality of our data can be deduced from the analysis of Ref.\
\cite{saoPauloTool}. We calculate a set of resonance-averaged
$S(E)$-factors. Their values are uncertain due to nuclear physics
effects -- due to using the S\~ao Paulo model with the NL3 nucleon
density distribution. The uncertainties were estimated
\cite{saoPauloTool} in comparison with other theoretical models and
available experimental data. For the reactions involving stable
nuclei, typical uncertainties are expected to be within a factor of
2, with maximum up to a factor of 4. For the reactions involving
unstable nuclei, typical uncertainties can be as large as one order
of magnitude, reaching two orders of magnitude at low energies for
the reactions with very neutron-rich isotopes. These uncertainties
reflect the current state of art in modelling the nuclear structure
and fusion reaction mechanism; they affect $S(E)$ and should be
taken into account while using the data. The advantage of our data
set is that it is wide and uniform. Note that the uncertainties in
$S(E)$, a factor of 2--10, may have no strong effect on the results
of modelling of nuclear burning phenomena \cite{saoPauloTool}.

\section{Analytic approximation of the astrophysical S-factors}
\label{fits}

All $S$-factors calculated here have been approximated by the
analytic expression
\begin{equation}
  S(E) = \exp \left\{ B_1+B_2E + B_3 E^2
  + {C_1+C_2 E + C_3 E^2 +C_4 E^3 \over
  1+\exp \left[(E_C-E)/D \right ]} \right \}.
\label{sfit}
\end{equation}
In this expression, $E$ is a center-of-mass energy of reacting
nuclei expressed in MeV, and $E_C$, $D$; $B_1$, $B_2$, $B_3$;
$C_1$, $C_2$, $C_3$, and $C_4$ are nine fit parameters for each
reaction. These parameters are described below; their values are
given in Tables \ref{tab:cc}--\ref{tab:mgmg} -- one table for each
of the reaction types listed in Table \ref{tab:reactions}. We
express $S(E)$ in MeV~b.

Equation (\ref{sfit}) generalizes the expression we suggested
earlier (e.g., Ref.\ \cite{saoPauloTool}) with some terms rearranged
for easier use. It contains additional fit parameters to accommodate
a much larger number of fusion reactions and a broader energy range
with one format. The $S(E)$-data calculated for each reaction were
fitted using Eq.\ (\ref{sfit}) over a total energy range (from 2~MeV
to $E_\mathrm{max}$) indicated in Table \ref{tab:reactions}.

The fit parameters in Eq.\ (\ref{sfit}), presented in Tables
\ref{tab:cc}--\ref{tab:mgmg} for different reactions, have clear
physical meaning:
\begin{itemize}
\item The parameter $E_C$ (expressed in MeV) is approximately
equal to the height of the Coulomb barrier. It divides the energy
range into that below the barrier ($E \lesssim E_C$) and that
above the barrier ($E \gtrsim E_C$), where $S(E)$-curves show
distinctly different behaviors.

\item
The parameter $D$ (expressed in MeV) characterizes a narrow energy
width ($D \sim 1$ MeV) of the transition region (at $|E- E_C|
\lesssim 2 D$) between the energy ranges below and above the Coulomb
barrier. This transition is governed by the function (of the
Fermi-Dirac type)
\begin{equation}
   {1 \over 1+\exp [(E_C-E)/D] },
\label{FD}
\end{equation}
which tends to 0 below the barrier and tends to 1 above the
barrier.

\item The parameter $B_1$ determines $S(0)$ (expressed in MeV~b):
\begin{equation}
   S(0)= \exp(B_1) \quad \mathrm{MeV~b}.
\label{S(0)}
\end{equation}
The accuracy of extrapolating the approximated $S(E)$ to $E \to 0$
is high, as discussed below.

\item The parameters $B_2$ (expressed in MeV$^{-1}$) and $B_3$
(expressed in MeV$^{-2}$) specify the energy dependence of $S(E)$
at $E \lesssim E_C$ as
\begin{equation}
   S(E)=S(0)  \exp (B_2 E+ B_3 E^2) \quad \mathrm{MeV~b}.
\label{below}
\end{equation}
Therefore, $S(E)$ at $E\lesssim E_C$ is actually determined by
\textit{three} parameters $B_1$, $B_2$, and $B_3$ out of the 9. This
sub-barrier energy range is sufficient for the majority of
applications to stellar burning (Sec.\ \ref{s:rates}). The
3-parameter fit remains quite accurate (within few tens percent) at
$E\leq E_C-2 D$ (but becomes divergent at higher $E$; see Sec.\
\ref{example} for an example). The same parameters $B_1$, $B_2$, and
$B_3$ contribute also to the $S(E)$ dependence at $E \gtrsim E_C$
[see Eq.\ (\ref{above})].

\item
The parameters $C_1$ (dimensionless), $C_2$ (expressed in
MeV$^{-1}$), $C_3$ (expressed in MeV$^{-2}$), and $C_4$ (expressed
in MeV$^{-3}$), together with  $B_1$, $B_2$, and $B_3$, specify the
$S(E)$ dependence at $E \gtrsim E_C$ (say, at $E \geq E_C+2D$),
\begin{equation}
    S(E)=S(0)\,\exp[C_1+(B_2+C_2)E+(C_3+B_3)E^2+C_4 E^3]
    \quad \mathrm{MeV~b}.
\label{above}
\end{equation}

\item In the last columns of Tables
\ref{tab:cc}--\ref{tab:mgmg} we give $\delta$ (in percentage), which
is the maximum relative error of fitted $S(E)$ values for each
reaction over the energy grid points.

\end{itemize}

Note that the maximum fit error $\delta$ does not exceed 16\% for
all the reactions in this study (and is much lower in many cases).
The maximum errors $\sim$10--16\% occur only for reactions involving
very neutron-rich nuclei. For the majority of reactions, maximum
errors are realized either at $E \sim E_C$ or at $E \sim
E_\mathrm{max}$. Root-mean-square relative errors over the energy
grid points are a factor of 2 to 3 lower than maximum errors
$\delta$. Considering the much larger uncertainties associated with
the nuclear-physics input of the calculated $S(E)$ (Sec.\
\ref{theory}), our fit accuracy can be regarded as unnecessarily
good. We keep it because it is good for a wide uniform data set. We
expect that the same Eq.\ (\ref{sfit}) can be used to approximate
$S(E)$ for other reactions, as well as for the here discussed
reactions when recalculated (in the future) with more advanced
nuclear physics models. As for the present $S(E)$ data, they could
have been approximated by Eq.\ (\ref{sfit}) retaining 7 parameters
out of the 9 (putting $B_3=C_4=0$). The fit errors would be higher;
in some cases the maximum errors would reach 30--40\%.

Our calculations of $S(E)$ are limited by $E \geq 2$ MeV. Because of
technical reasons (to solve the problem of quantum tunneling through
a very thick Coulomb barrier) we cannot directly calculate $S(0)$.
However, our fit expression (\ref{sfit}) is arranged in such a way
that it insures the low-energy dependence $S(E)=\exp(B_1+B_2 E+B_3
E^2)$ that is predicted on theoretical grounds (e.g., Ref.\
\cite{fh64}). Moreover, while fitting the data we have tried to
reach best fit accuracy at low $E$ (that is most important for
astrophysical applications). For that purpose, any $S(E)$ fit was
done in three stages. First, we have fitted all data points by Eq.\
(\ref{sfit}) and determine preliminary values of all fit parameters.
Second, we have fitted only subbarier $S(E)$ points (2 MeV $\leq E
\leq E_C-2.5D$) by Eq.\ (\ref{below}) to find $B_1$, $B_2$, and
$B_3$ (obtaining thus an accurate description at low $E$). Third, we
have refitted the entire $S(E)$ data set by Eq.\ (\ref{sfit}) with
the fixed values of $B_1$, $B_2$, and $B_3$ (as found at the
previous stage) and obtain thus final values of $E_C$, $D$,
$C_1$,\dots $C_4$. Accordingly, we expect that the fit enables one
to correctly determine the expansion terms $B_1$, $B_2$, and $B_3$,
and extrapolate to $E=0$. To check this point, we calculated
$S(1\,\mbox{MeV})=1.732 \times 10^{84}$ MeV~b for the
$^{46}$Mg+$^{46}$Mg reaction; it differs from the extrapolated value
of $1.839 \times 10^{84}$ MeV~b only by 6\%. This indicates that the
extrapolation to $E \to 0$ should be reliable.

\begin{figure}[tbh]
\begin{center}
\includegraphics[width=10.0cm,angle=0,bb=45 180 420 630]{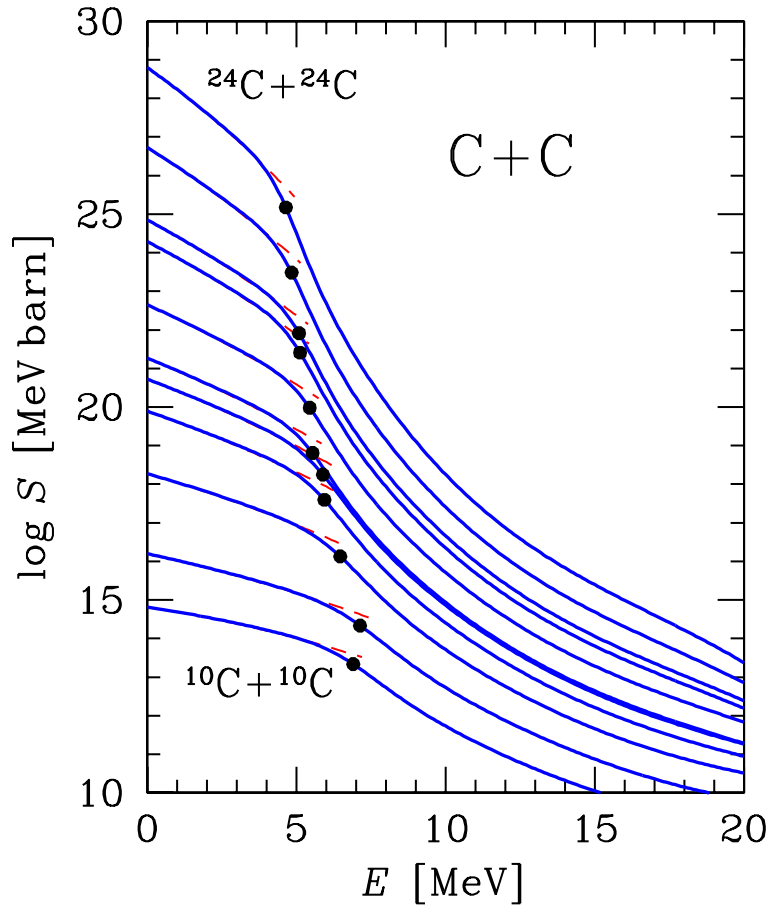}
\caption{(Color online) Analytic approximations of $S(E)$ for the
C+C reactions involving different isotopes. The curves from bottom
to top refer to ($A_1,A_2$)= (10,10), (12,12), (12,16), (12,20),
(16,16), (12,24), (16,20), (16,24), (20,20), (20,24), and (24,24)
reactions. Solid curves show the 9-parameter approximation
(\ref{sfit}); dashed curves (plotted at $E \leq E_C+0.3$ MeV) show
the 3-parameter approximation (\ref{below}). Filled dots correspond
to $E=E_C$. } \label{fig:cc}
\end{center}
\end{figure}

As an example, Fig.~\ref{fig:cc} presents the approximated $S(E)$
dependence for the C+C reactions with ($A_1,A_2$)= (10,10), (12,12),
(12,16), (12,20),  (16,16), (12,24), (16,20), (16,24), (20,20),
(20,24), and (24,24). The solid lines show our 9-parameter
approximation (\ref{sfit}); they coincide with the calculated data
points in the adopted logarithmic $S$-scale. Thus we do not present
the calculated points to simplify the figure. We plot our fit
expression not only in the energy range, where calculations are done
(from 2 to 17.9 MeV, Table \ref{tab:reactions}), but extrapolate
also beyond this range (from $E=2$ MeV to $E=0$, and from 17.9 MeV
to 20 MeV).  The thick dots mark the Coulomb barrier threshold,
$E=E_C$. One can see that, indeed, the behavior of $S(E)$ below and
above the Coulomb barrier is distinctly different, and the
transition from one regime to the other takes place within a narrow
energy range at $E \approx E_C$. For heavier isotopes (with larger
nucleus radii), the Coulomb barrier decreases, which increases the
$S(E)$ curve. The value of $S(0)$ for the reaction involving
heaviest isotopes ($^{24}$C+$^{24}$C) is approximately 15 orders of
magnitude larger than for the reaction of lightest isotopes
($^{10}$C+$^{10}$C). Finally, the dashed curves in Fig.~\ref{fig:cc}
show our 3-parameter $S(E)$ approximation (\ref{below}). It is seen
to be fairly accurate below the Coulomb barrier, but becomes
inaccurate in the vicinity of $E=E_C$ and at larger $E$.

\begin{figure}[tbh]
\begin{center}
\includegraphics[width=10.0cm,angle=0,bb=45 180 420 630]{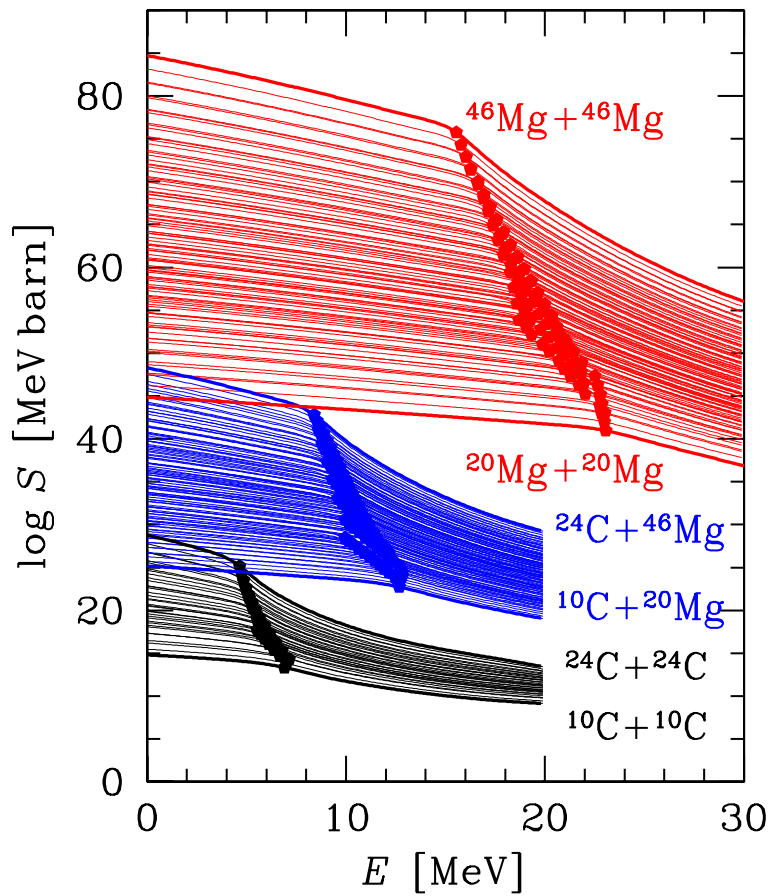}
\caption{(Color online) Analytic approximation (\ref{sfit}) of
$S(E)$ for 36 C+C reactions, 112 C+Mg reactions, and 105 Mg+Mg
reactions (Table \ref{tab:reactions}). The curves for each reaction
type are enclosed by the (thick) curves for the (indicated)
reactions involving lightest and heaviest isotopes. Filled dots
refer to $E=E_C$. } \label{fig:cmg}
\end{center}
\end{figure}

Figure \ref{fig:cmg} shows the $S(E)$ dependence for all 36 C+C
reactions, 112 C+Mg reactions, 105 Mg+Mg  fusion reactions
considered in this study (Table~\ref{tab:reactions}). The curves are
9-parameter fits (\ref{sfit}), and the thick dots again mark the
Coulomb barrier threshold, $E=E_C$. The curves for each fusion type
(C+C, C+Mg, Mg+Mg) are enclosed by the thick curves for the
reactions involving lightest and heavies isotopes (for instance,
$^{10}$C+$^{20}$Mg and  $^{24}$C+$^{46}$Mg, in the case of C+Mg
reactions). The general $S(E)$-behavior is seen to be the same as
for C+C reactions. The difference of $S(E)$ values, especially at
low $E$, for different reactions can be extremely large. For
example, the difference of $S(0)$ for the $^{46}$Mg+$^{46}$Mg and
$^{10}$C+$^{10}$C reactions is approximately 70 orders of magnitude.

\section{Numerical example}
\label{example}

\noindent As an example, we calculate $S(E)$ for the
$^{20}$Ne+$^{24}$Mg reaction at $E=$5 MeV. According to Table
\ref{tab:reactions}, fit parameters for the Ne+Mg reactions are
listed in Table \ref{tab:nemg}. The reaction in question corresponds
to $A_1=20$ and $A_2=24$. From line 17
of Table \ref{tab:nemg} we obtain the fit parameters:\\
$E_C=19.019$ MeV,
$D=0.88$ MeV;\\
$B_1=100.480$, $B_2=-0.3985$ MeV$^{-1}$,
$B_3=-0.00583$ MeV$^{-2}$;\\
$C_1=-40.079$, $C_2=6.6560$ MeV$^{-1}$, $C_3=-0.33863$ MeV$^{-2}$,
$C_4=0.005087$ MeV$^{-3}$.\\
Using Eq.\ (\ref{sfit}) at $E=5$ MeV we have
\begin{eqnarray}
  S(5~\mathrm{MeV}) & = & \exp \left\{
100.480-0.3985 \times 5 - 0.00583 \times 5^2
 {{}\over{}}   \right.
\nonumber \\
&+ & \left. {-40.079+6.6560\times 5  -0.33863 \times 5^2 + 0.005087
\times 5^3 \over
  1+\exp \left[(19.019-5)/0.88 \right ]} \right \}
\nonumber \\
& = & 5.1223 \times 10^{42}
   \quad \mathrm{MeV~b}.
\label{sfit1}
\end{eqnarray}
In this particular example, the fitted value deviates only by
0.006\% from the value
   $S(5~\mathrm{MeV})=5.1226$ MeV~b
calculated using the S\~ao Paulo potential.

\begin{figure}[t]
\begin{center}
\includegraphics[width=10.0cm,angle=0]{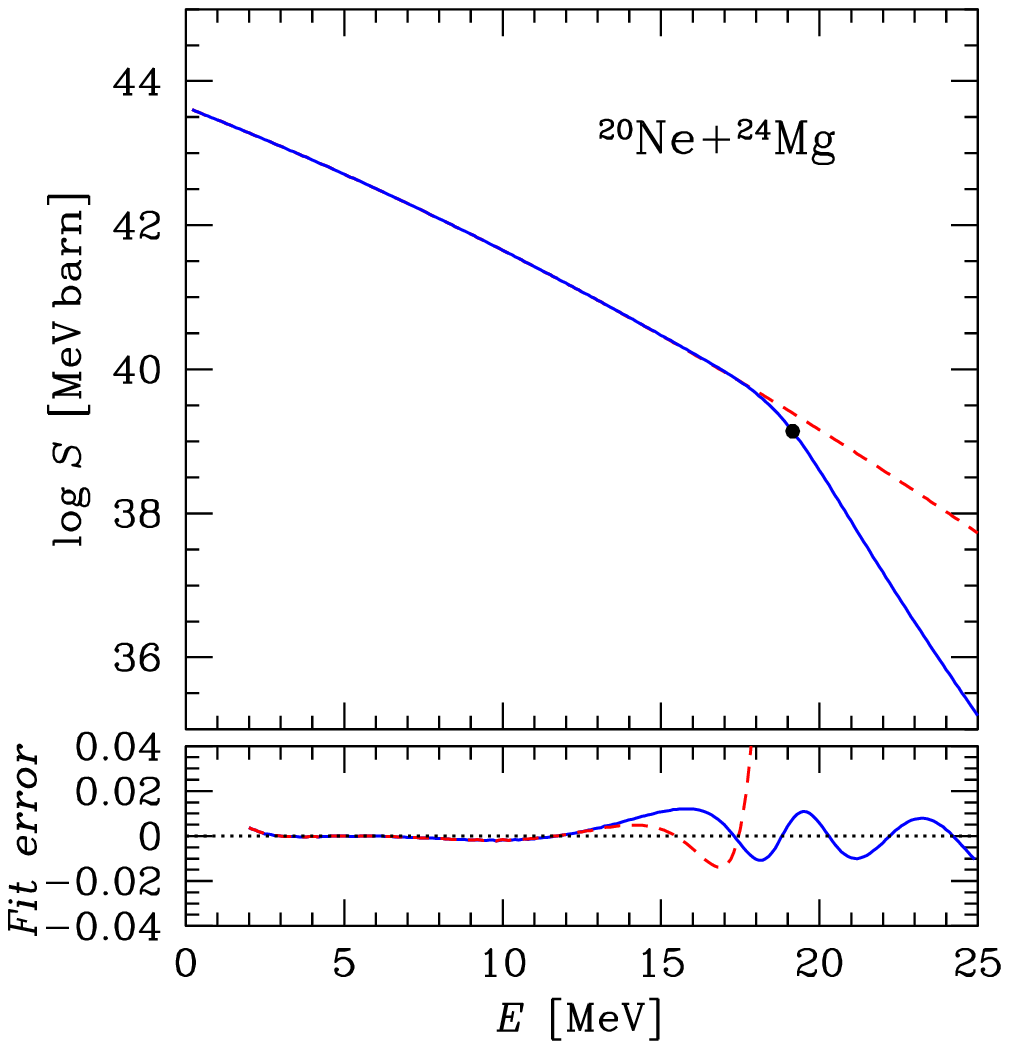}
\caption{(Color online) {\em Top}: Analytic approximations of $S(E)$
for the $^{20}$Ne+$^{24}$Mg reaction. The solid and dashed lines
are, respectively, the 9- and 3-parameter approximations
(\ref{sfit}) and (\ref{below}). The filled dot refers to $E=E_C$.
{\em Bottom}: Relative errors of these approximations with respect
to computed data. The dotted line shows zero error to guide the eye.
} \label{fig:nemg}
\end{center}
\end{figure}

Since our chosen energy $E=5$ MeV is lower than the Coulomb barrier
height, $E_C\approx 19$ MeV, we can also calculate $S(E)$ using our
3-parameter fit. From Eq.\ (\ref{S(0)}) we obtain
\begin{equation}
    S(0)=\exp(100.480)=4.3459 \times 10^{43}\quad \mathrm{MeV~b},
\label{S(0)1}
\end{equation}
and from Eq.\ (\ref{below}) we have
\begin{eqnarray}
     S(5~\mathrm{MeV}) & = & 4.3459 \times 10^{43}
  \exp (
       -0.3985 \times 5 - 0.00583 \times 5^2)
\nonumber \\
    & = & 5.1223 \times 10^{42} \quad \mathrm{MeV~b},
\label{below1}
\end{eqnarray}
in excellent agreement with the result (\ref{sfit1}) of the
9-parameter fit.

The $S(E)$ dependence for the $^{20}$Ne+$^{24}$Mg reaction is
plotted in the upper panel of Fig.~\ref{fig:nemg}. The solid and
dashed lines are the 9- and 3-parameter fits, respectively. In the
lower panel we show relative errors of fitted values of $S(E)$ [not
of $\log S(E)$] with respect to the calculated values. We see that
the 9-parameter fit is accurate over the entire energy range, in
which original $S(E)$ values have been calculated (Table
\ref{tab:reactions}). The maximum fit error of $\approx$1.2\% occurs
at $E=15.8$ MeV (and the root-mean-square fit error over all grid
points is $\approx 0.6$\%). The three-parameter fit (\ref{below})
stays highly accurate below the Coulomb barrier but diverges when
$E$ exceeds $E_C$. For instance, at $E=25$ MeV it overestimates
$S(E)$ by more than two orders of magnitude.

\section{Reaction rates}
\label{s:rates}

In the following we discuss the calculation of fusion reaction rates
$R$ [cm$^{-3}$~s$^{-1}$] in stellar environments. Any reaction rate
$R$ depends on the environmental conditions such as temperature,
density and composition of stellar matter. We cannot present tables
of the reaction rates covering the entire parameter space but give a
short description how the tabulated $S$-factors can be easily
utilized for deriving the reaction rates.

We limit our discussion to the formalism for non-resonant reaction
rates since the calculated $S$-factors presented here are entirely
characterized by non-resonant cross section behavior. If the cross
section is dominated by strong resonances the formalism should be
modified by adding the resonance contribution separately (e.g.,
Ref.\ \cite{itohetal03}).

As a rule, the main contribution to non-resonant rates comes from
nucleus-nucleus collisions in a narrow energy range $E\approx
E_\mathrm{pk}$, and $S(E)$ is a slowly varying function of $E$. Then
$R$ can be expressed through $S(E_\mathrm{pk})$. In particular, in
the thermonuclear regime, neglecting the effects of plasma
screening, $E_\mathrm{pk}$ is the standard Gamow-peak energy and the
reaction rates are given by the well-known classical theory of
thermonuclear burning (e.g., Refs.\ \cite{bbfh57,fh64,clayton83}).
This regime is realized at sufficiently high temperatures of stellar
matter when the nuclei form almost ideal Boltzmann gas. Even in this
regime $E_\mathrm{pk}$ is typically lower than $E_C$, and $S(E)$ can
be approximated by our 3-parameter fit (\ref{below}) without any
loss of accuracy in the reaction rate. If, however, one needs an
accurate reaction rate at
so
high temperatures
that $E_\mathrm{pk} \gtrsim E_C$, one should exactly calculate
the rate by integrating over $E$ with the full 9-parameter fit
(\ref{sfit}) for $S(E)$. The same fit can also be used for
evaluating the reaction cross sections $\sigma(E)$ in astrophysical
studies and nuclear physics laboratory experiments.

At lower temperatures the Gamow peak approach remains as a valid
approximation
but one should take into account the
plasma screening effects and a possible transition to the
pycnonuclear burning regime. In this case the equations for reaction
rates should be modified. These modifications are described in the
literature (e.g., Refs.\ \cite{gas2005,yak2006,cdi07,cd09} and
references therein). They are model-dependent, but in any case they
contain the values of $S(E)$ at certain ``Gamow-peak'' energies
$E_\mathrm{pk}$ (that are also modified by the plasma screening and
pycnonuclear burning effects, and depend on $T$ and $\rho$). Such
energies are typically much lower than $E_C$, so that our
3-parameter fits apply.

Specifically, Ref.\ \cite{gas2005} gives the expressions
(Sec.~III.G of \cite{gas2005}) for reaction rates in one-component
ion plasma in all 5 burning regimes. Ref.\ \cite{yak2006}
generalizes these expressions (Sec.\ III.G of \cite{yak2006}) to
the case of multicomponent ion plasma. The authors of Refs.\
\cite{cdi07} and \cite{cd09} present more accurate calculations
and approximations of reaction rates in one-component and
multi-component ion plasma, respectively. They use the WKB Coulomb
tunneling approximation in a radial mean field plasma potential
(that was extracted from extended Monte Carlo simulations of
classical strongly coupled ion plasmas). Their results are valid
in the thermonuclear burning regime and in the intermediate
thermo-pycno nuclear regime (it is possible that they can also be
extended to lower temperatures). In these regimes, the enhancement
factors of nuclear reaction rates due to plasma screening effects,
given in Refs.\ \cite{gas2005,yak2006,cdi07,cd09}, are in a
reasonably good agreement. In the pycnonuclear regimes (at zero
temperature and with thermal enhancement) the results
\cite{gas2005,yak2006} are model dependent due to plasma physics
uncertainties. For that reason, the authors of
\cite{gas2005,yak2006} present optimal (recommended) as well as
maximum and minimum theoretical reaction rates.

Note an omission in Ref.\ \cite{gas2005}: the expression for the
parameter $\lambda$, given by Eq.~(24) in \cite{gas2005}, for
pycnonuclear burning models in face-centered cubic crystals has to
be divided by $2^{1/3}$. This omission did not affect our choice of
optimal, maximum, and minimum reaction rates in \cite{gas2005}, and
our results in \cite{yak2006}. In addition, notice two typos in
\cite{yak2006}. In Eq.\ (32) of \cite{yak2006} there should be the
exponent sign $\exp$ before the last term in brackets in the
expression for $F_\mathrm{pyc}$; in Eq.\ (33) the product $x_i x_j$
must be replaced by $X_i X_j$. Also notice that in Refs.\
\cite{gas2005,yak2006} we neglected the Coulomb shifts of nucleus
energy levels in dense matter in the expression for $E_\mathrm{pk}$
at low temperatures (in pycnonuclear burning regimes). These effects
can influence the reaction rates at low $T$ if $S(E)$ is not a very
slowly varying function of $E$. Such shifts of $E_\mathrm{pk}$ have
been mentioned in the literature (e.g., Ref.\ \cite{itohetal03}) and
were parameterized in \cite{cd09}. We recommend to use Eq.\ (35) of
Ref.\ \cite{cd09} to calculate $E_\mathrm{pk}$ at all densities and
temperatures. The use of Eq.\ (40) of Ref.\ \cite{yak2006} is also
possible (gives correct rates of the reactions of study, within
nuclear physics uncertainties, for those densities and temperatures,
where these reactions are most efficient).

\begin{figure}[t]
\begin{center}
\includegraphics[width=10.0cm,angle=0]{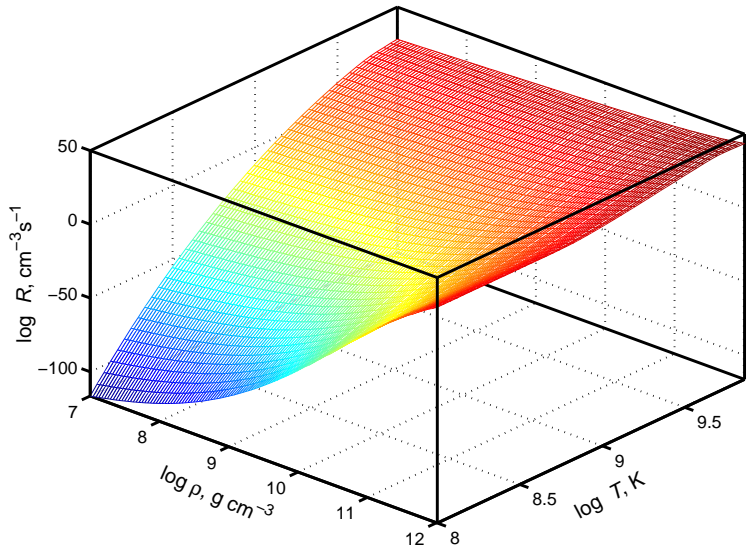}
\caption{(Color online) Density-temperature dependence of the
$^{20}$Ne+$^{24}$Mg reaction rate in an $^{20}$Ne-$^{24}$Mg mixture
with equal number fractions of Ne and Mg. } \label{fig:nemg1}
\end{center}
\end{figure}

For illustration of the procedure, Fig.\ \ref{fig:nemg1} presents
the rate of the previously discussed example $^{20}$Ne+$^{24}$Mg
as a function of density and temperature of matter composed of
$^{20}$Ne and $^{24}$Mg nuclei with equal number densities of Ne
and Mg nuclei. The rate is calculated using the optimal model of
the reaction rate from Ref.\ \cite{yak2006}. We show
%
a
wider $\rho-T$ range than the range, where these nuclei can really
exist in stellar matter, in order to demonstrate all features of the
reaction rate; these features are qualitatively the same for all
reactions of our study. One can see a strong temperature dependence
of the rate at high $T$ in the thermonuclear burning regimes, and a
strong density dependence
%
%
at high $\rho$ and low $T$ in the pycnonuclear regimes.

Finally, let us note again that nuclear physics uncertainties of our
calculated $S(E)$ are not small (Sec.\ \ref{theory}) and they
translate into uncertainties in the reaction rates. However, as we
analyzed in \cite{saoPauloTool}, such uncertainties are often not
very important for astrophysical applications.

\section{Conclusions}
\label{s:concl}

Using S\~ao Paulo method and the barrier penetration model we have
calculated the astrophysical $S$-factor as a function of energy for
946 fusion reactions involving various isotopes of C, O, Ne, and Mg,
from the stability valley to very neutron-rich nuclei. The
calculations have been performed on a dense grid of center-of-mass
energies $E$, from 2 MeV to 18--30 MeV, covering wide energy ranges
below and above the Coulomb barrier.

We fit calculated $S(E)$ values by a simple and accurate universal
9-parameter analytic formula, and present tables of fit parameters
for all the reactions. The fit error does not exceed 16\%. The
formula allows extrapolating $S(E)$ outside the considered energy
range, particularly, to lower energies $E \to 0$ of astrophysical
interest. We have also shown that the reduced fit expression,
containing 3 fit parameters out of 9, is fairly accurate at energies
below the Coulomb barrier and sufficient for accurate calculations
of the reaction rates at not too high temperatures of stellar
matter.

Our results can be used in computer codes for calculating nuclear
fusion rates and simulating various phenomena associated with
nuclear burning in high temperature and/or high density
astrophysical and laboratory plasmas. In particular, they can be
used for modelling nuclear burning in massive accreting white dwarfs
(type Ia supernovae) or in accreting neutron stars (superbursts,
deep crustal burning in X-ray transient sources). Nuclear burning at
high densities in these compact stars can involve neutron-rich
nuclei considered in the present paper.

The calculated $S(E)$ factors are rapidly varying functions of
energy $E$. Their values for various reactions can differ by many
orders of magnitude. On the other hand, their energy dependence
looks self-similar over a broad spectrum of the reactions. We will
address these findings in a separate publication.

\begin{acknowledgments}

DY is grateful to Andrew Chugunov for critical remarks and to Peter
Shternin for his assistance in the artwork. This work was partly
supported by the Joint Institute for Nuclear Astrophysics
(NSF-PHY-0822648), the U.S. Department of Energy under the grant
DE-FG02-07ER41459, the Russian Foundation for Basic Research (grants
08-02-00837 and 09-02-1208), and by the State Program ``Leading
Scientific Schools of Russian Federation'' (Grant NSh 2600.2008.2).
\end{acknowledgments}

\newpage

\begin{longtable}[c]{*{10}{c@{\extracolsep{\fill}}}c}
\caption{Fit parameters for C+C reactions} \label{tab:cc} \\
\hline
\hline 
$A_1$ $A_2$ & $E_C$ & $D$ & $B_1$ &$B_2$ & $B_3$ &
$C_1$ & $C_2$ & $C_3$ & $C_4$ & $\delta$\\
\hline \endfirsthead
\caption{Parameters for C+C reactions (continued)} 
\label{tab:cc} \\
\hline
\hline 
$A_1$ $A_2$ & $E_C$ & $D$ & $B_1$ &$B_2$ & $B_3$ &
$C_1$ & $C_2$ & $C_3$ & $C_4$ & $\delta$\\
\hline \endhead
  10  10 &
   6.871 &
    0.91 &
  34.213 &
  --0.3307 &
   --0.0109 &
   4.668 &
  --1.1271 &
    0.0348 &
    0.00021 &
      4 \\
  10  12 &
   6.988 &
    0.92 &
  35.713 &
  --0.3693 &
   --0.0121 &
   4.734 &
  --1.1528 &
    0.0378 &
    0.00018 &
      3 \\
  10  14 &
   6.663 &
    0.87 &
  38.031 &
  --0.4182 &
   --0.0166 &
   5.762 &
  --1.5058 &
    0.0646 &
   --0.00025 &
      4 \\
  10  16 &
   6.379 &
    0.83 &
  39.941 &
  --0.4733 &
   --0.0166 &
   6.550 &
  --1.7790 &
    0.0833 &
   --0.00062 &
      4 \\
  10  18 &
   6.123 &
    0.78 &
  41.647 &
  --0.5158 &
   --0.0183 &
   7.396 &
  --2.0754 &
    0.1051 &
   --0.00105 &
      5 \\
  10  20 &
   5.905 &
    0.73 &
  43.156 &
  --0.5415 &
   --0.0227 &
   8.261 &
  --2.3877 &
    0.1307 &
   --0.00150 &
      5 \\
  10  22 &
   5.718 &
    0.72 &
  44.608 &
  --0.6266 &
   --0.0169 &
   8.415 &
  --2.5025 &
    0.1368 &
   --0.00176 &
      6 \\
  10  24 &
   5.532 &
    0.68 &
  45.942 &
  --0.6489 &
   --0.0232 &
   9.160 &
  --2.8043 &
    0.1645 &
   --0.00224 &
      6 \\
\end{longtable}

\newpage

\begin{longtable}[c]{*{10}{c@{\extracolsep{\fill}}}c}
\caption{Fit parameters for C+O reactions} 
\label{tab:co} \\
\hline
\hline 
$A_{\rm C}$ $A_{\rm O}$ & $E_C$ & $D$ & $B_1$ &$B_2$ & $B_3$ &
$C_1$ & $C_2$ & $C_3$ & $C_4$ & $\delta$\\
\hline \endfirsthead
\caption{C+O reactions (continued)} 
\label{tab:co} \\
\hline
\hline 
$A_{\rm C}$ $A_{\rm O}$ & $E_C$ & $D$ & $B_1$ &$B_2$ & $B_3$ &
$C_1$ & $C_2$ & $C_3$ & $C_4$ & $\delta$\\
\hline \endhead 
  10  12 &
   8.852 &
    1.01 &
  41.846 &
  --0.3041 &
  --0.00687 &
   7.253 &
  --1.2828 &
   0.03300 &
  --0.000026 &
      4 \\
  10  14 &
   8.987 &
    1.03 &
  43.349 &
  --0.3366 &
  --0.00878 &
   7.175 &
  --1.2850 &
   0.03506 &
  --0.000019 &
      4 \\
  10  16 &
   8.846 &
    0.99 &
  45.102 &
  --0.3716 &
  --0.01071 &
   7.277 &
  --1.3364 &
   0.03825 &
  --0.000013 &
      4 \\
  10  18 &
   8.633 &
    0.98 &
  46.767 &
  --0.4065 &
  --0.01172 &
   7.928 &
  --1.4902 &
   0.04489 &
  --0.000014 &
      4 \\
  10  20 &
   8.411 &
    0.94 &
  48.281 &
  --0.4381 &
  --0.01273 &
   8.381 &
  --1.6072 &
   0.04990 &
  --0.000013 &
      5 \\
  10  22 &
   8.170 &
    0.90 &
  49.719 &
  --0.4684 &
  --0.01350 &
   8.855 &
  --1.7305 &
   0.05484 &
  --0.000008 &
      5 \\
  10  24 &
   7.896 &
    0.84 &
  51.164 &
  --0.5000 &
  --0.01397 &
   9.395 &
  --1.8751 &
   0.06053 &
  --0.000012 &
      5 \\
  10  26 &
   7.683 &
    0.81 &
  52.507 &
  --0.5320 &
  --0.01509 &
   9.691 &
  --1.9846 &
   0.06568 &
  --0.000007 &
      5 \\
\end{longtable}

\newpage

\begin{longtable}[c]{*{10}{c@{\extracolsep{\fill}}}c}
\caption{Fit parameters for C+Ne reactions} 
\label{tab:cne} \\
\hline
\hline 
$A_{\rm C}$ $A_{\rm Ne}$ & $E_C$ & $D$ & $B_1$ &$B_2$ & $B_3$ &
$C_1$ & $C_2$ & $C_3$ & $C_4$ & $\delta$\\
\hline \endfirsthead
\caption{C+Ne reactions (continued)} 
\label{tab:cne} \\
\hline
\hline 
$A_{\rm C}$ $A_{\rm Ne}$ & $E_C$ & $D$ & $B_1$ &$B_2$ & $B_3$ &
$C_1$ & $C_2$ & $C_3$ & $C_4$ & $\delta$\\
\hline \endhead
  10  18 &
  10.785 &
    1.04 &
  51.730 &
  --0.3382 &
  --0.00756 &
   7.339 &
  --0.8795 &
  --0.00206 &
   0.000700 &
      4 \\
  10  20 &
  10.645 &
    1.02 &
  53.219 &
  --0.3648 &
  --0.00852 &
   7.479 &
  --0.9120 &
  --0.00205 &
   0.000775 &
      4 \\
  10  22 &
  10.478 &
    1.00 &
  54.584 &
  --0.3888 &
  --0.00931 &
   7.652 &
  --0.9445 &
  --0.00261 &
   0.000865 &
      4 \\
  10  24 &
  10.293 &
    0.97 &
  55.866 &
  --0.4115 &
  --0.00995 &
   7.929 &
  --0.9977 &
  --0.00218 &
   0.000935 &
      4 \\
  10  26 &
  10.000 &
    0.92 &
  57.301 &
  --0.4362 &
  --0.01044 &
   8.333 &
  --1.0738 &
  --0.00214 &
   0.001043 &
      4 \\
  10  28 &
   9.774 &
    0.89 &
  58.622 &
  --0.4612 &
  --0.01129 &
   8.548 &
  --1.1307 &
  --0.00184 &
   0.001138 &
      4 \\
  10  30 &
   9.577 &
    0.86 &
  59.832 &
  --0.4838 &
  --0.01213 &
   8.712 &
  --1.1802 &
  --0.00170 &
   0.001231 &
      5 \\
  10  32 &
   9.303 &
    0.83 &
  61.127 &
  --0.5103 &
  --0.01195 &
  10.051 &
  --1.4841 &
   0.01442 &
   0.000957 &
      5 \\
\end{longtable}

\newpage

\begin{longtable}[c]{*{10}{c@{\extracolsep{\fill}}}c}
\caption{Fit parameters for C+Mg reactions} 
\label{tab:cmg} \\
\hline
\hline 
$A_{\rm C}$ $A_{\rm Mg}$ & $E_C$ & $D$ & $B_1$ &$B_2$ & $B_3$ &
$C_1$ & $C_2$ & $C_3$ & $C_4$ & $\delta$\\
\hline \endfirsthead
\caption{C+Mg reactions (continued)} 
\label{tab:cmg} \\
\hline
\hline 
$A_{\rm C}$ $A_{\rm Mg}$ & $E_C$ & $D$ & $B_1$ &$B_2$ & $B_3$ &
$C_1$ & $C_2$ & $C_3$ & $C_4$ & $\delta$\\
\hline \endhead
  10  20 &
  12.671 &
    1.03 &
  57.977 &
  --0.3116 &
  --0.00573 &
   8.521 &
  --0.8419 &
  --0.00042 &
   0.000348 &
      4 \\
  10  22 &
  12.658 &
    1.04 &
  59.259 &
  --0.3332 &
  --0.00662 &
   8.438 &
  --0.8461 &
  --0.00031 &
   0.000385 &
      4 \\
  10  24 &
  12.575 &
    1.04 &
  60.466 &
  --0.3529 &
  --0.00717 &
   8.456 &
  --0.8540 &
  --0.00113 &
   0.000443 &
      4 \\
  10  26 &
  12.472 &
    1.03 &
  61.545 &
  --0.3696 &
  --0.00769 &
   8.689 &
  --0.8855 &
  --0.00122 &
   0.000494 &
      4 \\
  10  28 &
  12.149 &
    1.01 &
  63.014 &
  --0.3908 &
  --0.00823 &
   9.285 &
  --0.9701 &
  --0.00127 &
   0.000595 &
      4 \\
  10  30 &
  11.904 &
    0.99 &
  64.303 &
  --0.4114 &
  --0.00885 &
   9.742 &
  --1.0452 &
  --0.00052 &
   0.000665 &
      4 \\
  10  32 &
  11.680 &
    0.98 &
  65.484 &
  --0.4315 &
  --0.00935 &
  10.131 &
  --1.1089 &
  --0.00021 &
   0.000740 &
      4 \\
  10  34 &
  11.441 &
    0.95 &
  66.631 &
  --0.4505 &
  --0.00984 &
  10.552 &
  --1.1750 &
  --0.00019 &
   0.000826 &
      4 \\
\end{longtable}

\newpage

\begin{longtable}[c]{*{10}{c@{\extracolsep{\fill}}}c}
\caption{Fit parameters for O+O reactions} \label{tab:oo} \\
\hline
\hline 
$A_1$ $A_2$ & $E_C$ & $D$ & $B_1$ &$B_2$ & $B_3$ &
$C_1$ & $C_2$ & $C_3$ & $C_4$ & $\delta$\\
\hline \endfirsthead
\caption{O+O reactions (continued)} 
\label{tab:oo} \\
\hline
\hline 
$A_1$ $A_2$ & $E_C$ & $D$ & $B_1$ &$B_2$ & $B_3$ &
$C_1$ & $C_2$ & $C_3$ & $C_4$ & $\delta$\\
\hline \endhead
  12  12 &
  11.072 &
    0.88 &
  51.217 &
  --0.2806 &
  --0.00380 &
   1.695 &
   0.5056 &
  --0.09860 &
   0.002627 &
      2 \\
  12  14 &
  11.158 &
    0.88 &
  53.344 &
  --0.3124 &
  --0.00535 &
   1.115 &
   0.6232 &
  --0.10635 &
   0.002850 &
      3 \\
  12  16 &
  11.015 &
    0.87 &
  55.697 &
  --0.3492 &
  --0.00650 &
   1.594 &
   0.5337 &
  --0.10435 &
   0.002914 &
      4 \\
  12  18 &
  10.801 &
    0.84 &
  57.885 &
  --0.3811 &
  --0.00741 &
   2.629 &
   0.3424 &
  --0.09724 &
   0.002883 &
      4 \\
  12  20 &
  10.584 &
    0.84 &
  59.873 &
  --0.4113 &
  --0.00822 &
   4.984 &
  --0.1294 &
  --0.07106 &
   0.002433 &
      4 \\
  12  22 &
  10.387 &
    0.89 &
  61.742 &
  --0.4396 &
  --0.00892 &
   9.920 &
  --1.1790 &
  --0.00443 &
   0.001069 &
      5 \\
  12  24 &
  10.069 &
    0.81 &
  63.594 &
  --0.4699 &
  --0.00926 &
   8.430 &
  --0.8352 &
  --0.03529 &
   0.001891 &
      5 \\
  12  26 &
   9.834 &
    0.80 &
  65.324 &
  --0.4999 &
  --0.01014 &
  10.098 &
  --1.2103 &
  --0.01413 &
   0.001538 &
      4 \\
\end{longtable}

\newpage

\begin{longtable}[c]{*{10}{c@{\extracolsep{\fill}}}c}
\caption{Fit parameters for O+Ne reactions} 
\label{tab:one} \\
\hline
\hline 
$A_{\rm O}$ $A_{\rm Ne}$ & $E_C$ & $D$ & $B_1$ &$B_2$ & $B_3$ &
$C_1$ & $C_2$ & $C_3$ & $C_4$ & $\delta$\\
\hline \endfirsthead
\caption{O+Ne reactions (continued)} 
\label{tab:one} \\
\hline
\hline 
$A_{\rm O}$ $A_{\rm Ne}$ & $E_C$ & $D$ & $B_1$ &$B_2$ & $B_3$ &
$C_1$ & $C_2$ & $C_3$ & $C_4$ & $\delta$\\
\hline \endhead
  12  18 &
  13.663 &
    1.00 &
  63.903 &
  --0.3116 &
  --0.00485 &
   5.015 &
   0.0871 &
  --0.06101 &
   0.001498 &
      3 \\
  12  20 &
  13.483 &
    0.99 &
  65.909 &
  --0.3379 &
  --0.00553 &
   5.219 &
   0.0748 &
  --0.06376 &
   0.001619 &
      3 \\
  12  22 &
  13.278 &
    0.97 &
  67.739 &
  --0.3618 &
  --0.00608 &
   5.427 &
   0.0692 &
  --0.06740 &
   0.001761 &
      4 \\
  12  24 &
  13.053 &
    0.94 &
  69.446 &
  --0.3843 &
  --0.00653 &
   5.518 &
   0.0883 &
  --0.07290 &
   0.001943 &
      4 \\
  12  26 &
  12.709 &
    0.90 &
  71.291 &
  --0.4078 &
  --0.00692 &
   5.756 &
   0.0991 &
  --0.08028 &
   0.002194 &
      4 \\
  12  28 &
  12.440 &
    0.87 &
  73.004 &
  --0.4324 &
  --0.00746 &
   5.936 &
   0.0890 &
  --0.08480 &
   0.002378 &
      4 \\
  12  30 &
  12.202 &
    0.84 &
  74.580 &
  --0.4558 &
  --0.00795 &
   5.819 &
   0.1282 &
  --0.09188 &
   0.002607 &
      5 \\
  12  32 &
  11.886 &
    0.82 &
  76.212 &
  --0.4775 &
  --0.00813 &
   8.572 &
  --0.3496 &
  --0.06986 &
   0.002284 &
      5 \\
\end{longtable}

\newpage

\begin{longtable}[c]{*{10}{c@{\extracolsep{\fill}}}c}
\caption{Fit parameters for O+Mg reactions} 
\label{tab:omg} \\
\hline
\hline 
$A_{\rm O}$ $A_{\rm Mg}$ & $E_C$ & $D$ & $B_1$ &$B_2$ & $B_3$ &
$C_1$ & $C_2$ & $C_3$ & $C_4$ & $\delta$\\
\hline \endfirsthead
\caption{O+Mg reactions (continued)} 
\label{tab:omg} \\
\hline
\hline 
$A_{\rm O}$ $A_{\rm Mg}$ & $E_C$ & $D$ & $B_1$ &$B_2$ & $B_3$ &
$C_1$ & $C_2$ & $C_3$ & $C_4$ & $\delta$\\
\hline \endhead
  12  20 &
  15.538 &
    0.82 &
  71.663 &
  --0.2887 &
  --0.00339 &
 --34.164 &
   6.6247 &
  --0.39665 &
   0.007019 &
      1 \\
  12  22 &
  15.453 &
    0.80 &
  73.440 &
  --0.3096 &
  --0.00407 &
 --36.470 &
   7.0372 &
  --0.42159 &
   0.007519 &
      2 \\
  12  24 &
  15.730 &
    0.91 &
  75.096 &
  --0.3293 &
  --0.00448 &
  --8.701 &
   2.4216 &
  --0.17185 &
   0.003078 &
      2 \\
  12  26 &
  15.587 &
    0.90 &
  76.580 &
  --0.3467 &
  --0.00483 &
  --9.511 &
   2.6103 &
  --0.18582 &
   0.003386 &
      2 \\
  12  28 &
  15.226 &
    0.88 &
  78.468 &
  --0.3669 &
  --0.00520 &
  --8.490 &
   2.5666 &
  --0.19241 &
   0.003634 &
      2 \\
  12  30 &
  14.944 &
    0.87 &
  80.145 &
  --0.3862 &
  --0.00567 &
  --7.412 &
   2.4775 &
  --0.19466 &
   0.003786 &
      2 \\
  12  32 &
  14.708 &
    0.86 &
  81.688 &
  --0.4055 &
  --0.00605 &
  --5.251 &
   2.1815 &
  --0.18432 &
   0.003693 &
      3 \\
  12  34 &
  14.427 &
    0.84 &
  83.174 &
  --0.4239 &
  --0.00637 &
  --5.525 &
   2.3143 &
  --0.19873 &
   0.004066 &
      3 \\
\end{longtable}

\newpage

\begin{longtable}[c]{*{10}{c@{\extracolsep{\fill}}}c}
\caption{Fit parameters for Ne+Ne reactions} \label{tab:nene} \\
\hline
\hline 
$A_1$ $A_2$ & $E_0$ & $D$ & $B_1$ &$B_2$ & $B_3$ &
$C_1$ & $C_2$ & $C_3$ & $C_4$ & $\delta$\\
\hline \endfirsthead
\caption{Ne+Ne reactions (continued)} 
\label{tab:nene} \\
\hline
\hline 
$A_1$ $A_2$ & $E_0$ & $D$ & $B_1$ &$B_2$ & $B_3$ &
$C_1$ & $C_2$ & $C_3$ & $C_4$ & $\delta$\\
\hline \endhead
  18  18 &
  16.681 &
    0.90 &
  81.201 &
  --0.3503 &
  --0.00538 &
   0.409 &
   0.8146 &
  --0.07429 &
   0.001168 &
      3 \\
  18  20 &
  16.513 &
    0.91 &
  84.124 &
  --0.3790 &
  --0.00607 &
   0.633 &
   0.8438 &
  --0.08058 &
   0.001349 &
      3 \\
  18  22 &
  16.292 &
    0.90 &
  86.808 &
  --0.4049 &
  --0.00668 &
   0.815 &
   0.9120 &
  --0.09050 &
   0.001613 &
      3 \\
  18  24 &
  16.029 &
    0.89 &
  89.323 &
  --0.4298 &
  --0.00715 &
   0.029 &
   1.1468 &
  --0.11006 &
   0.002054 &
      4 \\
  18  26 &
  15.607 &
    0.88 &
  92.021 &
  --0.4565 &
  --0.00751 &
   1.014 &
   1.1522 &
  --0.12151 &
   0.002407 &
      3 \\
  18  28 &
  15.122 &
    0.80 &
  94.527 &
  --0.4838 &
  --0.00807 &
 --12.920 &
   3.6521 &
  --0.26984 &
   0.005248 &
      2 \\
  18  30 &
  14.863 &
    0.81 &
  96.844 &
  --0.5102 &
  --0.00854 &
  --5.166 &
   2.4635 &
  --0.21343 &
   0.004364 &
      3 \\
  18  32 &
  14.455 &
    0.79 &
  99.245 &
  --0.5366 &
  --0.00860 &
   0.537 &
   1.6675 &
  --0.18219 &
   0.003973 &
      4 \\
\end{longtable}

\newpage

\begin{longtable}[c]{*{10}{c@{\extracolsep{\fill}}}c}
\caption{Fit parameters for Ne+Mg reactions} 
\label{tab:nemg} \\
\hline
\hline 
$A_{\rm Ne}$ $A_{\rm Mg}$ & $E_C$ & $D$ & $B_1$ &$B_2$ & $B_3$ &
$C_1$ & $C_2$ & $C_3$ & $C_4$ & $\delta$\\
\hline \endfirsthead
\caption{Ne+Mg reactions (continued)} 
\label{tab:nemg} \\
\hline
\hline 
$A_{\rm Ne}$ $A_{\rm Mg}$ & $E_C$ & $D$ & $B_1$ &$B_2$ & $B_3$ &
$C_1$ & $C_2$ & $C_3$ & $C_4$ & $\delta$\\
\hline \endhead
  18  20 &
  19.641 &
    0.93 &
  91.571 &
  --0.3251 &
  --0.00401 &
   0.723 &
   0.7764 &
  --0.05783 &
   0.000679 &
      3 \\
  18  22 &
  19.633 &
    0.93 &
  94.202 &
  --0.3481 &
  --0.00466 &
   0.729 &
   0.7445 &
  --0.05590 &
   0.000652 &
      3 \\
  18  24 &
  19.521 &
    0.93 &
  96.663 &
  --0.3690 &
  --0.00512 &
   0.618 &
   0.8060 &
  --0.06152 &
   0.000775 &
      3 \\
  18  26 &
  19.388 &
    0.93 &
  98.889 &
  --0.3882 &
  --0.00548 &
   0.653 &
   0.8649 &
  --0.06761 &
   0.000910 &
      3 \\
  18  28 &
  18.921 &
    0.93 &
 101.648 &
  --0.4108 &
  --0.00587 &
   0.770 &
   1.0376 &
  --0.08553 &
   0.001299 &
      4 \\
  18  30 &
  18.567 &
    0.93 &
 104.113 &
  --0.4328 &
  --0.00631 &
   0.576 &
   1.1900 &
  --0.09999 &
   0.001614 &
      4 \\
  18  32 &
  18.241 &
    0.92 &
 106.388 &
  --0.4545 &
  --0.00668 &
   0.546 &
   1.3077 &
  --0.11239 &
   0.001893 &
      4 \\
  18  34 &
  17.894 &
    0.91 &
 108.585 &
  --0.4754 &
  --0.00701 &
   0.587 &
   1.4214 &
  --0.12515 &
   0.002186 &
      4 \\
\end{longtable}

\newpage

\begin{longtable}[c]{*{10}{c@{\extracolsep{\fill}}}c}
\caption{Fit parameters for Mg+Mg reactions} \label{tab:mgmg} \\
\hline
\hline 
$A_1$ $A_2$ & $E_C$ & $D$ & $B_1$ &$B_2$ & $B_3$ &
$C_1$ & $C_2$ & $C_3$ & $C_4$ & $\delta$\\
\hline \endfirsthead
\caption{Mg+Mg reactions (continued)} 
\label{tab:mgmg} \\
\hline
\hline 
$A_1$ $A_2$ & $E_C$ & $D$ & $B_1$ &$B_2$ & $B_3$ &
$C_1$ & $C_2$ & $C_3$ & $C_4$ & $\delta$\\
\hline \endhead
  20  20 &
  23.050 &
    0.99 &
 103.307 &
  --0.2980 &
  --0.00312 &
 --27.040 &
   4.2173 &
  --0.18663 &
   0.002291 &
      4 \\
  20  22 &
  22.995 &
    0.99 &
 106.454 &
  --0.3206 &
  --0.00363 &
 --27.727 &
   4.3219 &
  --0.19218 &
   0.002384 &
      3 \\
  20  24 &
  22.844 &
    0.99 &
 109.389 &
  --0.3417 &
  --0.00396 &
 --30.112 &
   4.6524 &
  --0.20753 &
   0.002608 &
      2 \\
  20  26 &
  22.752 &
    0.99 &
 112.050 &
  --0.3614 &
  --0.00420 &
 --21.394 &
   3.6841 &
  --0.17273 &
   0.002191 &
      2 \\
  20  28 &
  22.031 &
    0.91 &
 115.256 &
  --0.3836 &
  --0.00447 &
 --50.742 &
   7.3809 &
  --0.32658 &
   0.004271 &
      2 \\
  20  30 &
  21.706 &
    0.92 &
 118.137 &
  --0.4044 &
  --0.00484 &
 --36.203 &
   5.8094 &
  --0.27250 &
   0.003654 &
      2 \\
  20  32 &
  21.305 &
    0.88 &
 120.802 &
  --0.4250 &
  --0.00515 &
 --38.554 &
   6.2535 &
  --0.29765 &
   0.004066 &
      2 \\
  20  34 &
  20.967 &
    0.88 &
 123.368 &
  --0.4450 &
  --0.00541 &
 --30.687 &
   5.4368 &
  --0.27208 &
   0.003808 &
      3 \\
\end{longtable}





\end{document}